# The miscalibration of the honeybee odometer

Laura Luebbert and Lior Pachter
Division of Biology and Biological Engineering
California Institute of Technology

We examine a series of articles on honeybee odometry and navigation published between 1996 and 2010, and find inconsistencies in results, duplicated figures, indications of data manipulation, and incorrect calculations. This suggests that redoing the experiments in question is warranted.

**Introduction**

When foraging honeybees return to the colony, they perform a dance that has baffled scientists since the time of Aristotle more than 2,300 years ago (Aristotle, 384–322 BC). The meaning of the honeybee dance was first decoded by Karl von Frisch almost a hundred years ago in his book "Aus dem Leben der Bienen" (Frisch, 1927), a discovery for which he won the Nobel Prize in 1973. Von Frisch discovered that honeybees use dance to convey information about the location and distance of food sources. He showed that different types of dances, so-called round and waggle dances, convey information about food sources that are less or more than 100 meters away, respectively. Von Frisch also found that the angle of the dancing bee in relation to gravity corresponds to the angle of the direction of the food in relation to the sun. However, the physiology underlying how bees estimate their traveled distance to the food source, as well as the encoding of the dances into exact distances, remained elusive.

**The miscalibration of the honeybee odometer**

In the "Honeybee Navigation: Nature and Calibration of the 'Odometer'" (Srinivasan *et al.*, 2000), the authors describe experiments to investigate how honeybees estimate and communicate the distance to food sources. By placing food at the end of tunnels lined with visual cues and subsequently recording the waggle dance performed by foraging bees, the authors provide evidence that distance estimation is visually driven. Based on estimations of the image motion experienced by the bees in the tunnels, the authors moreover calculate that each millisecond of waggle dance encodes 17.7° of image motion. The authors "propose that this is the fundamental, absolute calibration of the honeybee's visual odometer." The 17.7° per millisecond of waggle dance measure is widely cited in the field of entomology and insect-inspired robotics, see for example Egelhaaf and Kern (2002), Labhart and Meyer (2002), Cheng (2012), and Esch (2012).

According to Srinivasan *et al.* (2000), the 17.7° was calculated "from the slope of the regression line in Fig. 2" from which the authors "calculate that 186 m of outdoor flight is encoded by a waggle duration of 350 ms." We have reproduced Figure 2 (Srinivasan *et al.*, 2000) from the original paper in our Figure 1, with an additional red line at 186 m of outdoor flight and an orange line at 350 ms of waggle duration. We noticed that the data does not support that 186 m of outdoor flight is encoded by a waggle duration of 350 ms unless the y-intercept (96 ms) is ignored. As the authors stated, they used only the slope in their calculation. This approach does not produce a calibration from which absolute waggle duration and image motion may be deduced. When the intercept is taken into account, 186 m of outdoor flight corresponds to 445.86 ms of waggle duration. Hence,

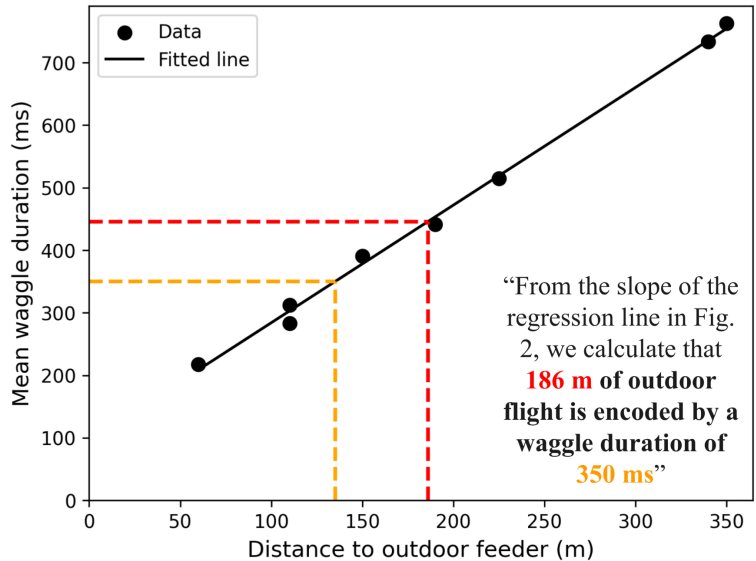

Figure 2 from M. V. Srinivasan et al. "Honeybee Navigation: Nature and Calibration of the 'Odometer'". *Science* (2000). DOI:10.1126/science.287.5454.851
**Inconsistencies between numbers reported in the text and those supported by the data**

"From the slope of the regression line in Fig. 2, we calculate that **186 m of outdoor flight is encoded by a waggle duration of 350 ms**"

**Figure 1:** Annotated distance-duration plot using the data from Table 1 / Figure 2 of Srinivasan *et al.* (2000).

following the author's reasoning, 1 ms of waggle dance should encode approximately 6180 / 445.86 = 13.86° of image motion in the eye. This revised number, i.e. 13.86°, is closer to the result of Esch (2012), who calibrated that a millisecond of waggle dance encodes 10.95° of image motion.

During the calculations above, we examined the data underlying Figure 2 of Srinivasan *et al.*, 2000, which we displayed in Table S1 here. According to the authors, waggle dances "were video-filmed at 25 frames per second and were later played back for frame-by-frame analysis. The duration of each waggle phase was measured in terms of the number of frames over which it occurred." This means that the waggle duration must have been measured in increments of 40 ms (1000 ms / 25 frames). As a result, we expect the reported means multiplied by the number of replicates to be divisible by 40. As shown in Table S1, this is not the case for any of the reported means and replicate numbers.

We next examined the data of Srinivasan *et al.* (2000) in the context of other studies aimed at decoding the relationship between distance and dance duration. Schürch *et al.* (2019) combine modern and historical data on the correlation between waggle duration and distance to compute a universal model that the authors claim "may be used, irrespective of landscape or subspecies." It is notable that the data from Srinivasan *et al.* (2000) has a 56% lower intercept (0.096 vs. 0.17) and 36% higher slope (1.88 vs. 1.38) in comparison to the universal calibration by Schürch *et al.* (2019) (Figure 2). This is unexpected as the different calibrations "differed significantly in intercept but not slope" (Schürch *et al.*, 2019).

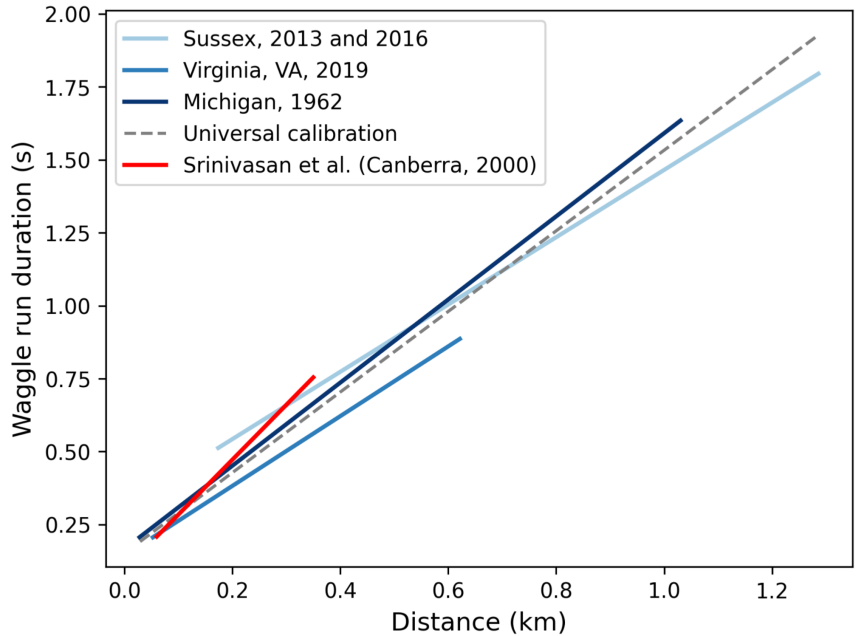

**Figure 2:** Data from Figure 2 of Schürch *et al.* (2019) and Figure 2 of Srinivasan *et al.* (2000) plotted together.

Based on the discrepancy between results obtained by Srinivasan *et al.* (2000) and other research groups, we also examined other data published between 1996 and 2010.

**Data duplication and potential manipulation**

In the 1997 paper "Visually Mediated Odometry in Honeybees" by M. V. Srinivasan, S. W. Zhang, and N. J. Bidwell, which examines different odometric cues potentially underlying distance estimation in bees, identical control graphs are (re)used across most panels in Figures 2, 3, and 5. We have reproduced these panels in Supp. Figure 1. While the 1997 paper does not state that most experiments did not include their own control in the experimental design, the control conditions (tunnel length = 3.2 m) and the number of replicates (N = 121) are reported consistently for these plots. However, a graph identical to the reused control graph appears again in Figure 7 of the 1997 paper, and in that figure, the reported number of replicates and experimental conditions are not concordant with those reported for identical data in the prior figures (reported N = 71 and a tunnel length of 7.6 m, Figure 3A).

Duplication of identical data with different reported experimental conditions occurs at least once more in the 1997 paper: Figures 3A and 3B of the original paper contain an identical graph, labeled in 3A with N=19 and experimental condition "still air" and in 3B with N=35 and labeled as control (see Figure 3B here).

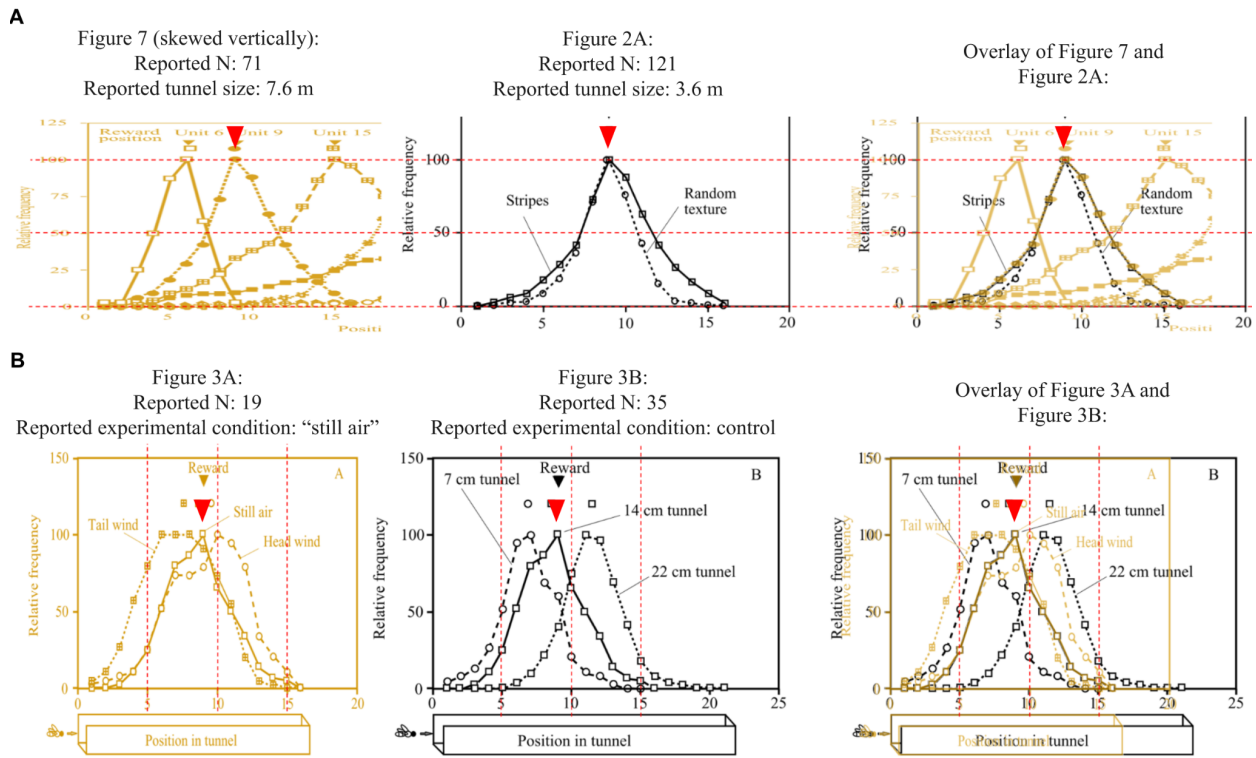

Figures from M. V. Srinivasan, S. W. Zhang, N. J. Bidwell; Visually Mediated Odometry in Honeybees. *J Exp Biol* 1 October 1997; 200 (19): 2513–2522. doi: https://doi.org/10.1242/jeb.200.19.2513
**Identical data reported with different experimental conditions**

A
Figure 7 (skewed vertically):
Reported N: 71
Reported tunnel size: 7.6 m

Figure 2A:
Reported N: 121
Reported tunnel size: 3.6 m

Overlay of Figure 7 and Figure 2A:

B
Figure 3A:
Reported N: 19
Reported experimental condition: "still air"

Figure 3B:
Reported N: 35
Reported experimental condition: control

Overlay of Figure 3A and Figure 3B:

**Figure 3:** Instances of identical data being reported for different experimental conditions in Srinivasan *et al.* (1997). These figures were reproduced from the original publication under a CC-BY license.

Several of the figures from the 1997 paper were first published in another paper, namely "Honeybee navigation en route to the goal: visual flight control and odometry" by M. V. Srinivasan, S. W. Zhang, M. Lehrer, and T. S. Collett (1996). We found the following inconsistencies between the two papers:

- All graphs shown in Figure 8 from the 1996 paper and Figure 3 of the 1997 paper are identical but are labeled with different experimental conditions (tunnel width of 11 cm in 1996 versus 7 cm in 1997) and different numbers of replicates (N=88 in 1996 and N=56 in 1997).

- The training tunnel used across experiments was described as having a length of 3.35 m in 1996 compared to 3.20 m in the 1997 paper.

- The spatial distributions of the honeybee search shown in Figure 6C in the 1996 paper and in Figure 1C in the 1997 paper differ, even though the rest of the data is identical.

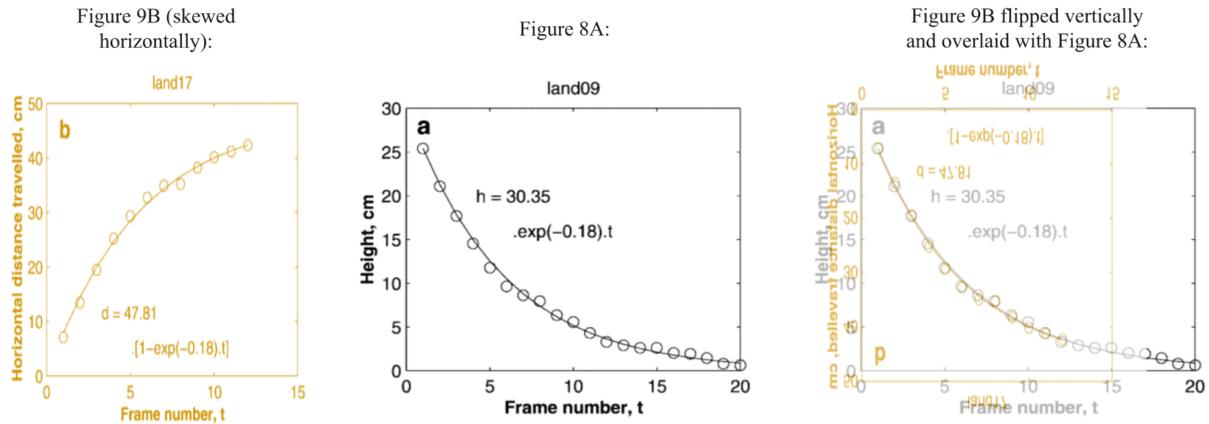

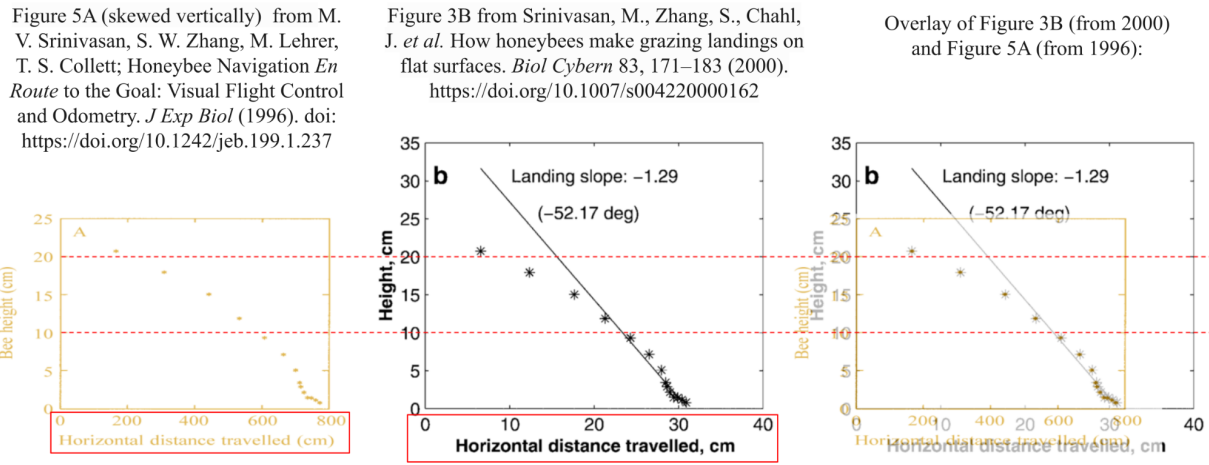

**Figure 4:** Instances of seemingly manipulated, highly similar or identical data reported for different experiments within the same and across different papers. These figures were reproduced from the original publications under a CC-BY license.

In addition to the repeated reporting of identical data for different experiments, we found several instances indicating the possibility of data manipulation: In the paper, "How honeybees make grazing landings on flat surfaces" by M. V. Srinivasan, S. W. Zhang, J. S. Chahl, E. Barth and S. Venkatesh (2000), the data in Figures 8A and 9B are highly similar, as evident when Figure 9B is slightly skewed and flipped vertically (Figure 4A). Furthermore, the first 6 measurements shown in Figures 9A, C, and D are identical (Supp. Figure 2, top left panel), the last 5 measurements shown in Figures 8A, B, and C are identical (Supp Figure 2, top right panel), and the first 5 measurements shown in Figures 8A and B are identical (Supp Figure 2, bottom left panel).

Finally, Figure 3B of the 2000 paper "How honeybees make grazing landings on flat surfaces"

displays data that is identical to that shown in Figure 5A of the 1996 paper "Honeybee Navigation En Route to the Goal: Visual Flight Control and Odometry", when Figure 5A is slightly skewed vertically (Figure 3B). The data is labeled with different x-axes in the two papers.

Above, we have documented several occurrences of data duplication across experimental conditions in honeybee navigation papers published between 1996 and 2000 (Figure 3, Supp. Figure 1). Moreover, as shown in Figure 4 and Supp. Figure 2, in some instances, the data appear to have been manipulated before duplication. Our findings are summarized in Table 1.

| Title | Year | Publisher | Authors | Issues | Associated figures (here) | DOI |
|---|---|---|---|---|---|---|
| Honeybee Navigation En Route to the Goal: Visual Flight Control and Odometry | 1996 | The Journal of Experimental Biology | M. V. Srinivasan, S. W. Zhang, M. Lehrer, T. S. Collett | Data duplicated within the same paper and between the 1996 and 1997 papers labeled with different experimental conditions | Figures 3 and 4B; Supp. Figure 1 | https://doi.org/10.1242/jeb.199.1.237 |
| Visually Mediated Odometry in Honeybees | 1997 | The Journal of Experimental Biology | M. V. Srinivasan, S. W. Zhang, N. J. Bidwell | Data duplicated within the same paper and between the 1996 and 1997 papers labeled with different experimental conditions | Figure 3; Supp. Figure 1 | https://doi.org/10.1242/jeb.200.19.2513 |
| How honeybees make grazing landings on flat surfaces | 2000 | Biological Cybernetics | M V Srinivasan, S W Zhang, J S Chahl, E Barth, S Venkatesh | Duplicated data from the 1996 paper (on one occasion with different axis); Several data duplications for different experimental conditions | Figure 4; Supp. Figure 2 | https://doi.org/10.1007/s004220000162 |
| Honeybee Navigation: Nature and Calibration of the 'Odometer' | 2000 | Science | Mandyam V. Srinivasan, Shaowu Zhang, Monika Altwein, Jürgen Tautz | Data inconsistent with reported methods and results from similar studies | Figures 1, 2, and 5; Table S1 | 10.1126/science.287.5454.851 |
| Landing Strategies in Honeybees and Applications to Uninhabited Airborne Vehicles | 2004 | The International Journal of Robotics Research | J. S. Chahl, M. V. Srinivasan, S. W. Zhang | Figures 11 and 12 are identical; Republishes problematic data from the 2000 Biological Cybernetics paper | / | 10.1177/02783649040413 20 |
| Visual Motor Computations in Insects | 2004 | Annual Review of Neuroscience | Mandyam V. Srinivasan and Shaowu Zhang | Republishes problematic data from the 2000 Biological Cybernetics paper and the 1997 paper | / | 10.1146/annurev.neuro.27.070203.144343 |

**Table 1:** Overview of papers with data duplication and potential manipulation issues.

**A correlation of 0.99 between $R^2 = 0.99$ regressions and honeybee navigation papers**

In examining the papers above, we noticed repeated presentations of regressions with $R^2 \geq 0.99$ (Table 2). These extremely high correlations suggest that across studies and experiments, there was very little technical and biological noise. Given that data was often measured across different hives, on variable terrain, and incorporating many different individual bees, this seemed unlikely (Towne and Gould, 1988; Beekman *et al.*, 2008; Okada *et al.*, 2014; Schürch *et al.*, 2016; Smith *et al.*, 2022). Schürch *et al.* (2016) show that the differences in dance duration-distance curves between individual bees are so high that differences are evident even when using limited frame rates.

First, we noticed that in Figure 1c of the 2005 paper "Visual working memory in decision making by honey bees" by S. Zhang, F. Bock, A. Si, J. Tautz, and M. V. Srinivasan, the authors report $R^2 = 0.999$. We analyzed the data from this figure and found $R^2 = 0.918$ (Figure 5). Zhang *et al.* report an incorrect coefficient of determination.

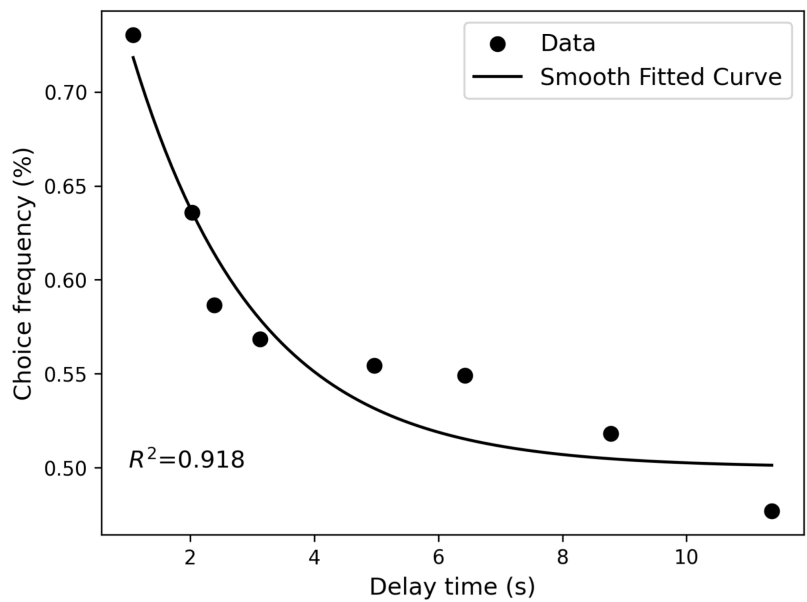

Figure 1c from S. Zhang *et al.* "Visual working memory in decision making by honey bees". *PNAS* (2005). DOI: 10.1073/pnas.0501440102
**Incorrect $R^2$ reported**

**Figure 5:** Choice frequency over delay time plot using data from Figure 1c from Zhang *et al.* (2005). Zhang *et al.* (2005) reported $R^2 = 0.999$. However, we find $R^2 = 0.918$.

Next, we examined the data by Srinivasan *et al.* (2000) that was collected using two separate bee hives. The authors claim that "No colony-specific or route-specific differences were apparent." However, with the exception of a single distance (110 m), different feeder distances were measured for the two hives. Hence, the data does not support the author's claim. The single feeder distance that was measured for both hives, 110 m, was encoded by an average 283.0 ± 66.0 ms of waggle dance by hive 1 and 312.4 ± 44.4 ms of waggle dance by hive 2, which is the greatest deviation from the regression line across all data points (Figure 1). It seems unlikely that all other mean dance durations for the different feeder distances fell almost exactly on the regression line, regardless of the hive.

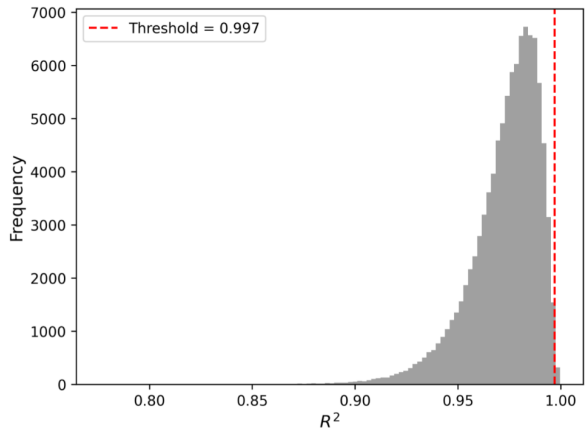
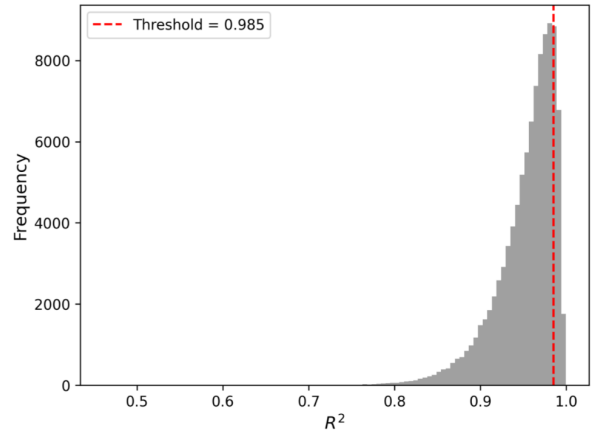

**Figure 6:** $R^2$ values computed 100,000 times by sampling from a Gaussian distribution with the reported mean and standard deviation values from Srinivasan *et al.* (2000) and Zhang *et al.* (2005).

To assess the likelihood of obtaining the reported $R^2$ values, we computed $R^2$ values by sampling from a Gaussian distribution with the reported mean and standard deviation values and computing the coefficient of determination $R^2$ 100,000 times for the Srinivasan *et al.* (2000) and Zhang *et al.* (2005) papers. While the mean values reported in Srinivasan *et al.* (2000) are almost perfectly linear with $R^2$ = 0.997, we found that only a fraction of 0.006 of the simulated $R^2$ values were equal to or greater than the reported $R^2$ value 0.997 (Figure 6, left panel).

In Srinivasan *et al.* (2000), the data was likely averaged over individual bees, dances, and waggle phases to generate a mean for each feeder location. In performing our analysis, we assumed that bees do not share the same mean (Schürch *et al.*, 2016). It is possible that averaging and binning, e.g. based on the camera frame rate, contributed to obtaining the high $R^2$ values that are reported throughout these papers. However, we took the camera frame rate into account in our analysis, and even when computing the $R^2$ values with the assumption that all bees and individual dances share the same mean, we obtained a fraction of 0.19 samples with $R^2 \geq 0.997$ (Supp. Figure 3).

Zhang *et al.* (2005) reported $R^2$ = 0.985 for their Figure 1b, a result that we were able to reproduce. However, we found that only a fraction of 0.14 of the simulated $R^2$ values using the data from Zhang *et al.* (2005) were greater than or equal to 0.985 (Figure 6, right panel). Taking into account both papers, we arrive at a probability of 0.19 * 0.14 = 0.0266 of both correlations being as high as reported. It would be highly unlikely to obtain all of the $R^2$ values shown in Table 2.

| Title | Year | Publisher | Authors | DOI | Reported $R^2$ |
|---|---|---|---|---|---|
| Honeybee Navigation: Nature and Calibration of the 'Odometer' | 2000 | Science | Mandyam V. Srinivasan, Shaowu Zhang, Monika Altwein, Jürgen Tautz | 10.1126/science.287.5454.851 | 0.997 |
| How honeybees make grazing landings on flat surfaces | 2000 | Biological Cybernetics | M V Srinivasan, S W Zhang, J S Chahl, E Barth, S Venkatesh | https://doi.org/10.1007/s004220000162 | 0.99 |
| Honeybee Odometry: Performance in Varying Natural Terrain | 2004 | PLOS Biology | Juergen Tautz, Shaowu Zhang, Johannes Spaethe, Axel Brockmann, Aung Si, Mandyam Srinivasan | https://doi.org/10.1371/journal.pbio.0020211 | 0.9899 |
| Visual working memory in decision making by honey bees | 2005 | PNAS | Shaowu Zhang, Fiola Bock, Aung Si, Juergen Tautz, Mandyam V. Srinivasan | https://doi.org/10.1073/pnas.0501440102 | 0.985 and 0.999 |
| Visual control of flight speed in honeybees | 2005 | The Journal of Experimental Biology | Emily Baird, Mandyam V. Srinivasan, Shaowu Zhang, Ann Cowling | https://doi.org/10.1242/jeb.01818 | 0.99 presented for Fig. 4 here: https://youtu.be/HJDFiuw9Djo?t=477 |
| The moment before touchdown: landing maneuvers of the honeybee Apis mellifera | 2010 | The Journal of Experimental Biology | C. Evangelista, P. Kraft, M. Dacke, J. Reinhard, M. V. Srinivasan | https://doi.org/10.1242/jeb.037465 | 0.99 |

**Table 2:** Overview of papers reporting $R^2 \geq 0.99$.

### Discussion

Our review argues for repeating the experiments in question, all published between 1996 and 2010, and verifying that the claims made are correct.

### Code availability

The code to reproduce the figures and analyses shown in this paper can be found here: https://github.com/pachterlab/LP_2024.

### Additional note

Some of the issues described here were first raised by Laura Luebbert and Elisabeth Bik in 2020 and can be found here:
https://pubpeer.com/publications/F639C63BF2C4AC4328070EEDA5857B
https://pubpeer.com/publications/9FE2AA69C8521CFEA14E453801350B

**Supplementary Data**

Examples of references to the 17.7° calibration of the honeybee visual odometer calculated by Srinivasan *et al.* (2000):

- Martin Egelhaaf, Roland Kern. Vision in flying insects. Current Opinion in Neurobiology, 2002 https://doi.org/10.1016/S0959-4388(02)00390-2:
  "At a mean distance between the honeybees and the tunnel wall of 5.5 cm, 1 ms of waggle in the dance corresponded to **17.7°** of image motion on the eyes."
- Thomas Labhart, Eric P. Meyer. Neural mechanisms in insect navigation: polarization compass and odometer. Current Opinion in Neurobiology, 2002 https://doi.org/10.1016/S0959-4388(02)00384-7:
  "[…] 1 ms of waggle duration corresponds to **17.7°** of image motion in the eye"
- Martin Egelhaaf. The neural computation of visual motion information. Invertebrate Vision, 2006:
  "At a mean distance of the honey bees to the tunnel wall of 5.5 cm, 1 ms of waggle in the dance corresponded to **17.7°** of image motion on the eyes (data redrawn from Srinivasan et al., 2000a)"
- Jonathan Peter. Behavioral and Theoretical Evidence that Non-directional Motion Detectors Underlie the Visual Estimation of Speed in Insects. The University of Arizona (2009):
  "[...] each 1 ms of the waggle dance signals **17.7 degrees** of image motion."
- Esch H. (2012) Foraging Honey Bees: How Foragers Determine and Transmit Information About Feeding Site Locations. In: Galizia C., Eisenhardt D., Giurfa M. (eds) Honeybee Neurobiology and Behavior. Springer, Dordrecht:
  "One millisecond of wagging required **17.7°** of image motion from front to back in the tunnel."
- Ken Cheng. How to navigate without maps: The power of taxon-like navigation in ants. Comparative Cognition & Behavior Reviews, 2012:
  "Srinivasan et al. (2000) have calibrated the waggle dance as indicating **~17.7°** of optic flow for each millisecond of waggling."
- Bergantin, Lucia. Robotic models for the honeybee visual odometer. Diss. Aix Marseille Université (AMU), 2023:
  "calibration for the honeybee visual odometer [...] equivalent to **17.7deg** of image motion per millisecond of waggle"

| Experiment | Mean waggle duration (ms) | n (number of bees ; number of dances ; number of waggle phases) | (Mean * n) / 40 for each n |
|---|---|---|---|
| Outdoor feeder at 60 m (H1) | 217.5 | 3 ; 10 ; 92 | 16.31 ; 54.38 ; 500.25 |
| Outdoor feeder at 110 m (H1) | 283.0 | 5 ; 10 ; 93 | 35.38 ; 70.75 ; 657.98 |
| Outdoor feeder at 110 m (H2) | 312.4 | 6 ; 14 ; 181 | 46.86 ; 109.34 ; 1413.61 |
| Outdoor feeder at 150 m (H1) | 390.9 | 3 ; 10 ; 92 | 29.32 ; 97.73 ; 899.07 |
| Outdoor feeder at 190 m (H1) | 441.2 | 6 ; 7 ; 65 | 66.18 ; 77.21 ; 716.95 |
| Outdoor feeder at 225 m (H2) | 514.4 | 8 ; 21 ; 345 | 102.88 ; 270.06 ; 4,436.7 |
| Outdoor feeder at 340 m (H2) | 733.3 | 10 ; 23 ; 222 | 183.33 ; 421.65 ; 4,069.82 |
| Outdoor feeder at 350 m (H1) | 762.6 | 5 ; 9 ; 87 | 95.33 ; 171.59 ; 1,658.66 |
| Tunnel experiment 2 | 528.8 | 4 ; 9 ; 216 | 52.88 ; 118.98 ; 2,855.52 |
| Tunnel experiment 4 | 441.2 | 7 ; 16 ; 138 | 77.21 ; 176.48 ; 1,522.14 |

**Table S1:** Mean and number of replicate (n) values reported for each experiment in "Honeybee Navigation: Nature and Calibration of the 'Odometer'" (Srinivasan *et al.*, 2000). The last column shows (mean * each reported type of n) / 40.

Figures from M. V. Srinivasan, S. W. Zhang, N. J. Bidwell; Visually Mediated Odometry in Honeybees. *J Exp Biol* 1 October 1997; 200 (19): 2513–2522. doi: https://doi.org/10.1242/jeb.200.19.2513

**Reuse of control data across experiments**

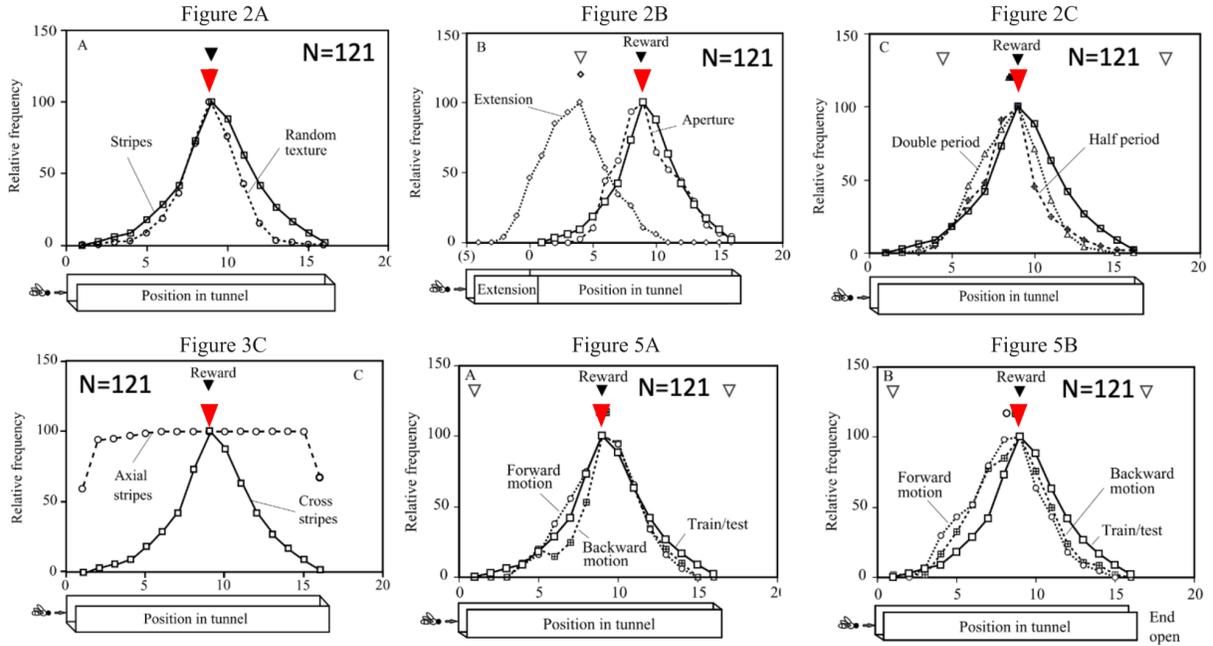

**Supplementary Figure 1:** Reuse of control data across different experiments in Srinivasan *et al.* (1997). These figures were reproduced from the original publication under a CC-BY license.

Figures from Srinivasan, M., Zhang, S., Chahl, J. *et al.* How honeybees make grazing landings on flat surfaces. *Biol Cybern* 83, 171–183 (2000). https://doi.org/10.1007/s004220000162

**Seemingly manipulated, partly identical data reported for different experiments**

Overlaid Figures 9A, C, and D:

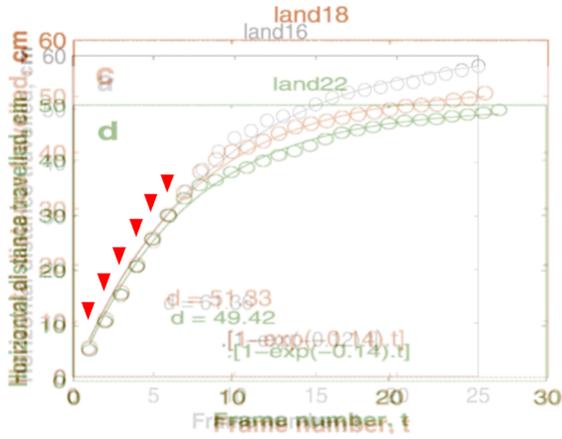

Overlaid Figures 8A, B, and C:

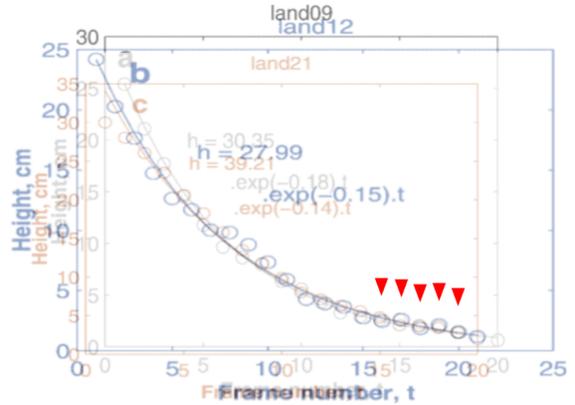

Overlaid Figures 8A and B:

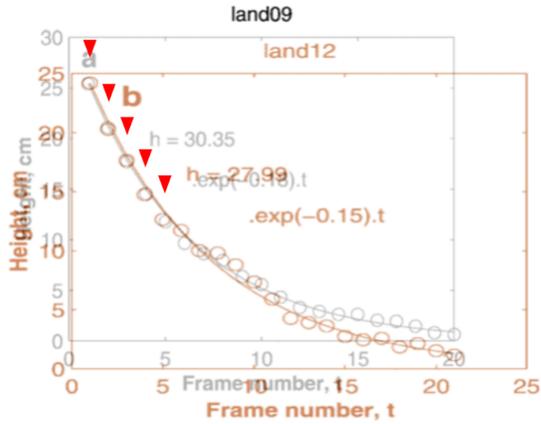

**Supplementary Figure 2:** Seemingly manipulated, partly identical data reported in Srinivasan *et al.* (2000). The red arrows mark potentially duplicated data points. These figures were reproduced from the original publication under a CC-BY license.

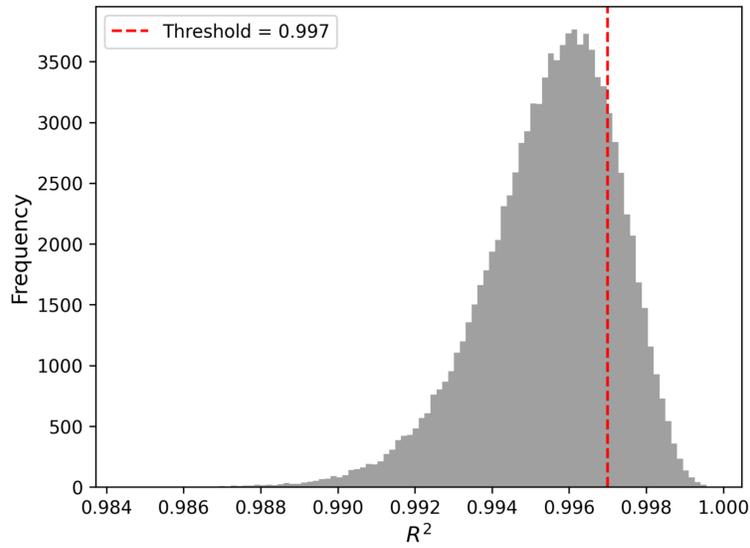

**Supplementary Figure 3:** $R^2$ values computed 100,000 times by sampling from a Gaussian distribution with the reported mean and standard deviation values from Srinivasan *et al.* (2000), assuming that individual bees and dances share the same mean.